\pdfoutput=1
\documentclass{PoS}

\usepackage{bm}
\usepackage{amsmath}

\newcommand{\tr}{\mathrm{tr}}
\newcommand{\I}{\mathrm{i}}
\def\Eq#1{Eq.~(\ref{#1})}
\def\Fig#1{Fig.~\ref{#1}}
\def\Sec#1{Sec.~\ref{#1}}
\def\Tab#1{Tab.~\ref{#1}}

\title{Strange nucleon form factors with $N_f = 2 + 1$ O($a$)-improved Wilson fermions}

\ShortTitle{Strange nucleon form factors}

\author{Dalibor Djukanovic$^{2}$, Harvey Meyer$^{1,2}$, Konstantin Ottnad$^{1,2}$, Georg von Hippel$^1$, \speaker{Jonas Wilhelm}$^1$, Hartmut Wittig$^{1,2}$\\
		\llap{$^1$}PRISMA Cluster of Excellence and Institute for Nuclear Physics\\        
        Johannes Gutenberg University of Mainz\\
		Johann-Joachim-Becher-Weg 45\\
		55128 Mainz\\
		\llap{$^2$}Helmholtz Institute Mainz\\
		Staudingerweg 18\\
		55128 Mainz\\
		\newline
        E-mail: {\email{D.Djukanovic@him.uni-mainz.de}, \email{meyerh@kph.uni-mainz.de}, \email{kottnad@uni-mainz.de}, \email{hippel@uni-mainz.de}, \email{jonas.wilhelm@uni-mainz.de}}, \email{wittig@kph.uni-mainz.de}}

\abstract{We present preliminary results for strangeness form factors of the nucleon computed on the CLS ensembles with $N_f=2+1$ flavors of O($a$)-improved Wilson fermions. Our calculations are performed at two values of the lattice spacing ($a \in \{0.064, 0.086\} \,\mathrm{fm}$) and three values of the pion mass ($m_\pi \in \{ 200, 280, 340\} \,\mathrm{MeV}$). The determination of strangeness form factors proceeds by computing quark-disconnected diagrams, for which we employ hierarchical probing in four dimensions in order to deal with this most challenging part of the calculation. Furthermore, we investigate several source-sink separations to check for excited-state contamination.}

\FullConference{The 36th Annual International Symposium on Lattice Field Theory - LATTICE2018\\
		22-28 July, 2018\\
		Michigan State University, East Lansing, Michigan, USA.}

\begin{document}

\section{Introduction}
Nucleon form factors are key quantities in hadron physics as they give us information about the internal structure of the nucleon. More specifically, the interaction of a nucleon with external currents acquires a momentum-transfer dependence, described by the form factors, because the nucleon is not a point-like particle. In this work, we focus on the axial form factor $G_A(Q^2)$ and the induced pseudoscalar form factor $G_P(Q^2)$, which parameterize the nucleon matrix element of the axial vector current
\begin{equation}
\left< N,\bm{p}^\prime,s^\prime \left| A_\mu(x) \right| N,\bm{p},s \right> = \bar{u}^{s^\prime}(\bm{p}^\prime)\ \widetilde{A}_\mu(q)\ u^{s}(\bm{p}) e^{\I q\cdot x}\ , 
\end{equation}
\begin{equation}
\widetilde{A}_\mu(q) = \gamma_\mu\gamma_5 G_{\text{A}}(Q^2) + \gamma_5 \frac{q_\mu}{2m_N}G_{\text{P}}(Q^2)\ ,
\end{equation}
where $p$ and $p^\prime$ are the four-momenta of the initial and final nucleon, and $Q^2=-q^2=-(p^\prime-p)^2$. These form factors are not only accessible in experiments \cite{Bernard} but also from a first-principles calculation using Lattice QCD, which is our chosen approach. Lattice QCD is a powerful tool for form factor calculations as it allows disentangling contributions from different quark flavors. Here, we focus on the contributions of the u, d and s quarks corresponding to quark-disconnected diagrams. The techniques to study the connected contributions of the u and d valence quarks, the only contributions required for the iso-vector form factors of the nucleon, are already well-established in the Mainz Lattice group \cite{Capitani}. Combining the connected and disconnected contributions to the axial vector form factors will enable us to determine the weak neutral current (WNC) axial form factor $G_A^Z(Q^2)$, obtained at leading order from the iso-vector contribution $G_A(Q^2)$ and the strange-quark contribution $G_A^s(Q^2)$ using SU(3) flavor symmetry \cite{GAZ}: $G_A^Z(Q^2) = -G_A(Q^2) + G_A^s(Q^2)$. In addition, we will construct the flavor non-singlet induced pseudoscalar form factor $G_P^8(Q^2)= G_P^{u+d}(Q^2) - 2 G_P^s(Q^2)$. Here, the light-quark contribution $G_P^{u+d}(Q^2)$ contains connected and disconnected quark contributions whereas the strange-quark contribution $G_P^s(Q^2)$ is solely disconnected. These quantities are of importance since it has been seen that the WNC $G_A^Z(Q^2)$ gives a main contribution to $\nu p$ and $\bar{\nu}p$ differential cross sections \cite{WNC}, while $G_P^8(Q^2)$ can be used to obtain the $\eta$-nucleon coupling $g_{\eta NN}$, if the $\eta$ decay constant $f_\eta^8$ is known \cite{JerGA}. 

\section{Extracting form factors from Lattice QCD}\label{sec:method}
The starting point to extract form factors from Lattice QCD is the nucleon three-point function
\begin{equation}
C_{3,A_\mu}^N(\bm{q},z_0;\bm{p}^\prime,y_0;\Gamma_\nu) = \sum_{\bm{y},\bm{z}} e^{i\bm{q}\bm{z}}e^{-i\bm{p}^\prime\bm{y}}\ (\Gamma_\nu)_{\beta\alpha} \left\langle N_{\alpha}(\bm{y},y_0)A_\mu(\bm{z},z_0)\bar{N}_\beta(0)\right\rangle_G
\end{equation}
with a nucleon interpolator $N_\alpha(x)$, a flavor-diagonal axial vector current $A_\mu(x)$ and a projector $\Gamma_\nu$. For the projector we consider
\begin{equation}
\Gamma_0 = \frac{1}{2}(1+\gamma_0)\ ,\ \Gamma_i = \frac{1}{2}(1+\gamma_0)\ \I\gamma_5\gamma_i\ ,\ i\in\{1,2,3\}\ ,
\end{equation}
where $\Gamma_0$ projects the nucleon to the correct parity, and $\Gamma_i$ additionally polarizes the nucleon spin along the $i$-axis. Applying the spectral decomposition to the nucleon three-point function and only taking the ground-state into account, which means that $z_0,(y_0-z_0)\gg0$, one arrives at
\begin{align} 
C_{3,A_\mu}^N(\bm{q},z_0;\bm{p}^\prime,y_0;\Gamma_\nu) &= f(\bm{p}^\prime,\bm{q},y_0,z_0)\ T\left(\widetilde{A}_\mu,\Gamma_\nu,\bm{q},\bm{p}^\prime\right)\ .
\end{align}
The function $f$ contains nucleon overlap factors, time dependencies and kinematic factors. To eliminate the first two, we construct a ratio of nucleon three-point and two-point functions \cite{ffinlqcd}
\begin{equation}
R_{A_\mu}(\bm{q},z_0;\bm{p}^\prime,y_0;\Gamma_\nu) = \frac{C_{3,A_\mu}^N(\bm{q},z_0;\bm{p}^\prime,y_0;\Gamma_\nu)}{C_2^N(\bm{p}^\prime,y_0;\Gamma_0)}\sqrt{\frac{C_2^N(\bm{p}^\prime,y_0;\Gamma_0)\ C_2^N(\bm{p}^\prime,z_0;\Gamma_0)\ C_2^N(\bm{p}^\prime\text{-}\bm{q},y_0\text{-}z_0;\Gamma_0)}{C_2^N(\bm{p}^\prime\text{-}\bm{q},y_0;\Gamma_0)\ C_2^N(\bm{p}^\prime\text{-}\bm{q},z_0;\Gamma_0)\ C_2^N(\bm{p}^\prime,y_0\text{-}z_0;\Gamma_0)}}\ ,
\label{eq:ratio}
\end{equation}
so that the spectral decomposition of the ratio yields for the ground-state
\begin{equation}
R_{A_\mu}(\bm{q};\bm{p}^\prime;\Gamma_\nu) = \frac{1}{4\sqrt{(E_{\bm{p}^\prime-\bm{q}}+m_N)(E_{\bm{p}^\prime}+m_N)E_{\bm{p}^\prime}E_{\bm{p}^\prime-\bm{q}}}}\ T\left(\widetilde{A}_\mu,\Gamma,\bm{q},\bm{p}^\prime\right)\ ,
\label{eq:sdrat}
\end{equation}
\begin{equation}
T\left(\widetilde{A}_\mu,\Gamma_\nu,\bm{q},\bm{p}^\prime\right) = \tr\left[ \Gamma_\nu \left( E_{\bm{p}^\prime}\gamma_0 -i\bm{p}^\prime\bm{\gamma} + m_N \right)\ \widetilde{A}_\mu(\bm{q})\ \left( E_{\bm{p}^\prime-\bm{q}}\gamma_0 -i(\bm{p}^\prime-\bm{q})\bm{\gamma} + m_N \right) \right]\ .
\end{equation}
The function $T$ can be calculated for all combinations of a component of the axial vector current $A_\mu(x)$ and a component of the projector $\Gamma_\nu$. Each combination leads to a kinematic prefactor for the axial and induced pseudoscalar form factor $M^A_{\nu\mu},\,M^P_{\nu\mu}$. \Eq{eq:sdrat} then takes the form
\begin{equation}
R_{A_\mu}(\bm{q};\bm{p}^\prime;\Gamma_\nu) = M_{\nu\mu}^A(\bm{q},\bm{p}^\prime)\ G_A(Q^2) + M_{\nu\mu}^P(\bm{q},\bm{p}^\prime)\ G_P(Q^2)\ .
\end{equation}
The individual prefactors can be grouped to form a matrix $M$ for each combination of $\bm{q}$ and $\bm{p}^\prime$ that correspond to the same value of $Q^2$. Similarly, we can form a vector $\bm{R}$ from the data for the ratios. Now we can define the (generally overdetermined) system of equations
\begin{equation}
M\ \bm{G} = \bm{R},\ \ M = \left( \begin{array}{c}
M^A_1\\
\vdots\\
M^A_N\\
\end{array}\ \begin{array}{c}
M^P_1\\
\vdots\\
M^P_N\\
\end{array} \right),\ \ \bm{G} = 
\left(
\begin{array}{c}
G_A(Q^2)\\
G_P(Q^2)\\
\end{array}
\right),\ \ \bm{R} = 
\left(
\begin{array}{c}
R_1\\
\vdots\\
R_N\\
\end{array}
\right),
\label{eq:system}
\end{equation}
which connects our lattice results for the ratios on the right-hand side to the analytical expectation from the spectral decomposition on the left-hand side. It can be solved for the form factors by minimizing the least-squares function \cite{Capitani}
\begin{equation}
\chi^2 = (\bm{R}-M\bm{G})^T\ C^{-1}\ (\bm{R}-M\bm{G})\ ,
\end{equation}
where the covariance matrix $C$ is approximated from the lattice data of the ratios. Before we actually solve the system in \Eq{eq:system}, two steps are done to reduce the system size $N$ and increase the statistical precision. We first drop all non-contributing equations ($M^A = 0\ \&\ M^P = 0$) and then average equivalent contributions\footnote{Example of two equivalent contributions:\\
\begin{equation*}
\left.\begin{array}{c}
\Gamma_1,A_2,\bm{p}_a=\left(1\ 0\ 0\right)^T,\bm{q}_a=\left(0\ 1\ 0\right)^T,\bm{p}_a^\prime=\left(1\ 1\ 0\right)^T\\
\Gamma_3,A_2,\bm{p}_b=\left(0\ 0\ 1\right)^T,\bm{q}_b=\left(0\ 1\ 0\right)^T,\bm{p}_b^\prime=\left(0\ 1\ 1\right)^T
\end{array} \right\} \Rightarrow M_{12}^A(\bm{q}_a,\bm{p}_a^\prime) = M_{32}^A(\bm{q}_b,\bm{p}_b^\prime)\ \&\ M_{12}^P(\bm{q}_a,\bm{p}_a^\prime) = M_{32}^P(\bm{q}_b,\bm{p}_b^\prime)
\end{equation*}}.
For the number of independent equations over our considered range of $Q^2$ values we find: $N\in\left\{ 4,  5,  8,  9, 10, 11, 12, 13, 14, 18, 19, 21, 22, 25, 26, 28, 34 \right\}$. Note that we perform the averaging procedure already for the nucleon three-point functions, with the additional constraint that the momenta for the nucleon states at the source and the sink are related by spatial symmetry \cite{AV3pt}. Furthermore, we average the nucleon two-point functions over equivalent momentum classes. We then calculate the ratios from these averaged correlation functions. As the left-hand side of the system of equations corresponds to the ground-state contribution, we perform fits to the asymptotic behavior or employ the summation method (see e.g. \cite{ffinlqcd,gace}) to isolate the ground-state contribution, also in the lattice data, before solving for the form factors.

\section{Simulation}\label{sec:simulation}

\subsection{Ensembles}
In this work we use CLS $N_f = 2+1$ O($a$)-improved Wilson fermion ensembles \cite{CLS}. The gauge sector is described by the tree-level improved L\"uscher-Weisz gauge action. These ensembles have open boundary conditions in time to prevent the problem of topological freezing, and approach the physical values of the quark masses along a $\tr\ M = \text{const}$ trajectory, where $M$ is the quark mass matrix. The subset of ensembles and configurations we have processed for this project so far is shown in \Tab{tab:ensembles}. We employ the improved local axial vector current 
\begin{equation}
A_\mu^f(\bm{z},z_0)^{\text{Imp.}}=\bar{f}(\bm{z},z_0) \gamma_5\gamma_\mu f(\bm{z},z_0)+ac_A\ \partial_\mu \left(\bar{f}(\bm{z},z_0) \gamma_5 f(\bm{z},z_0)\right)\ ,
\end{equation}
where we distinguish between the light and the strange quarks, $f\in\left\{l,s\right\}$, as the up and down quarks are degenerate on our ensembles. A non-perturbative determination of the improvement coefficient $c_A$ has been done in \cite{alpha}. As motivated in the introduction, we focus on the disconnected contributions. For this we need the flavor-singlet renormalization constant $Z_A^0$, which has not been determined yet, and thus we present unrenormalized (bare) results in \Sec{sec:results}. The three-point function corresponding to the disconnected contribution factorizes into separate traces for the quark loop and the nucleon two-point function
\begin{equation}
C_{3,A_\mu}^{N,l/s}(\bm{q},z_0;\bm{p}^\prime,y_0;\Gamma_\nu) = \left\langle \mathcal{L}_{A_\mu}^{l/s}(\bm{q},z_0)\cdot \mathcal{C}_2^N(\bm{p}^\prime,y_0;\Gamma_\nu) \right\rangle_G\ .
\end{equation}
These are the two main building blocks, described in more detail in the next two subsections.

\renewcommand{\arraystretch}{1.2}
\begin{center}
\begin{table}[h]
\center
\begin{tabular}{l|ccccccccc}
	&$\beta$	&$a$ [fm] &$N_s^3\times N_t$	&$m_\pi$[MeV]	&$m_K$[MeV]	&$N_{\text{cfg}}$	&$N_{\text{meas}}$\\
\hline
H105	&3.40	&0.086 &$32^3\times 96$	&280	&460	&1020	&391680\\
\hline
N203	&3.55 &0.064	&$48^3\times 128$ &340	&440	&772	&345856\\
N200	&3.55 &0.064	&$48^3\times 128$	&280	&460	&856	&383488\\
D200	&3.55 &0.064	&$64^3\times 128$	&200	&480	&278	&124544

\end{tabular}
\caption{The processed gauge ensembles for this work. $N_{\text{cfg}}$ denotes the number of gauge configurations. The last column corresponds to the total number of measurements for the ratio in \Eq{eq:ratio}.}
\label{tab:ensembles}
\end{table}
\end{center}
\renewcommand{\arraystretch}{1}

\subsection{Nucleon two-point function}\label{sec:2pt}
The nucleon two-point function is given by
\begin{equation}
C_2^N(\bm{p}^\prime,y_0;\Gamma_\nu) = \sum_{\bm{y}\in\Lambda} e^{-i\bm{p}^\prime\bm{y}}\ (\Gamma_\nu)_{\beta\alpha} \left\langle N_{\alpha}(y)\bar{N}_\beta(0) \right\rangle\ ,
\end{equation}
\begin{equation}
N_{\alpha}(x) = \epsilon_{abc}\left( u_\beta^a(x)\ \left(C\gamma_5\right)_{\beta\gamma}\ d_\gamma^b(x) \right)\ u_\alpha^c(x)\ .
\end{equation}
All quark propagators have been Wuppertal smeared \cite{wuppertal} at the source and the sink. We employ the truncated solver method \cite{trunc1,trunc2} to increase the statistical precision of the nucleon two-point functions at moderate cost. The trick is to first obtain a biased estimate with a large number of low-precision solves for the quark propagator and then add a bias correction from a much smaller subset of high-precision solves. We placed the sources for the nucleon two-point functions on seven timeslices for each ensemble. The seven timeslices were evenly distributed around the middle of the time extent. These were separated by seven timeslices on which no sources were placed. The number of high-precision solves on each timeslice was $N_{\text{src}}^{HP}=1$, except for H105, where we used $N_{\text{src}}^{HP} = 4$. For all ensembles, the number of low-precision solves on each timeslice was $N_{\text{src}}^{LP}=32$. Both the forward and the backward-propagating nucleon two-point functions from all source positions were included, except for the first (last) timeslice on H105, where we omitted the backward (forward) propagation. This is due to the arising boundary effects as H105 has a smaller temporal lattice extent than the other three ensembles. 

\subsection{Quark loop}
The calculation of the quark loop requires an all-to-all propagator, which can be stochastically estimated with noise vectors $\eta$
\begin{equation}
L_{A_\mu}^{l/s}(\bm{q},z_0) = -\sum_{\bm{z}\in\Lambda} e^{i\bm{q}\cdot\bm{z}}\ \left<tr\left[S^{l/s}(z;z)\ \gamma_5\gamma_\mu\right]\right>_G
 = -\sum_{\bm{z}\in\Lambda} e^{i\bm{q}\cdot\bm{z}}\ \left<\eta^{\dagger}(z)\ \gamma_5\gamma_\mu\ s^{l/s}(z)\right>_{G,\eta}\ .
\end{equation}
Here we use hierarchical probing \cite{hp}, which augments the series of noise vectors $\eta_n$ by a set of Hadarmard vectors $h_n$, where each element of a noise vector is multiplied with the Hadamard vectors to obtain an improved estimate of the quark loop. We employ four-dimensional noise and Hadamard vectors and use two independent noise vectors with 512 Hadamard vectors each. Thus, we perform a total of 1024 inversions per gauge configuration and flavor for the quark loop calculation.

\section{Results}\label{sec:results}
\begin{center}
\begin{figure}[h]
\begin{minipage}{\textwidth}
   \includegraphics[scale=0.4225]{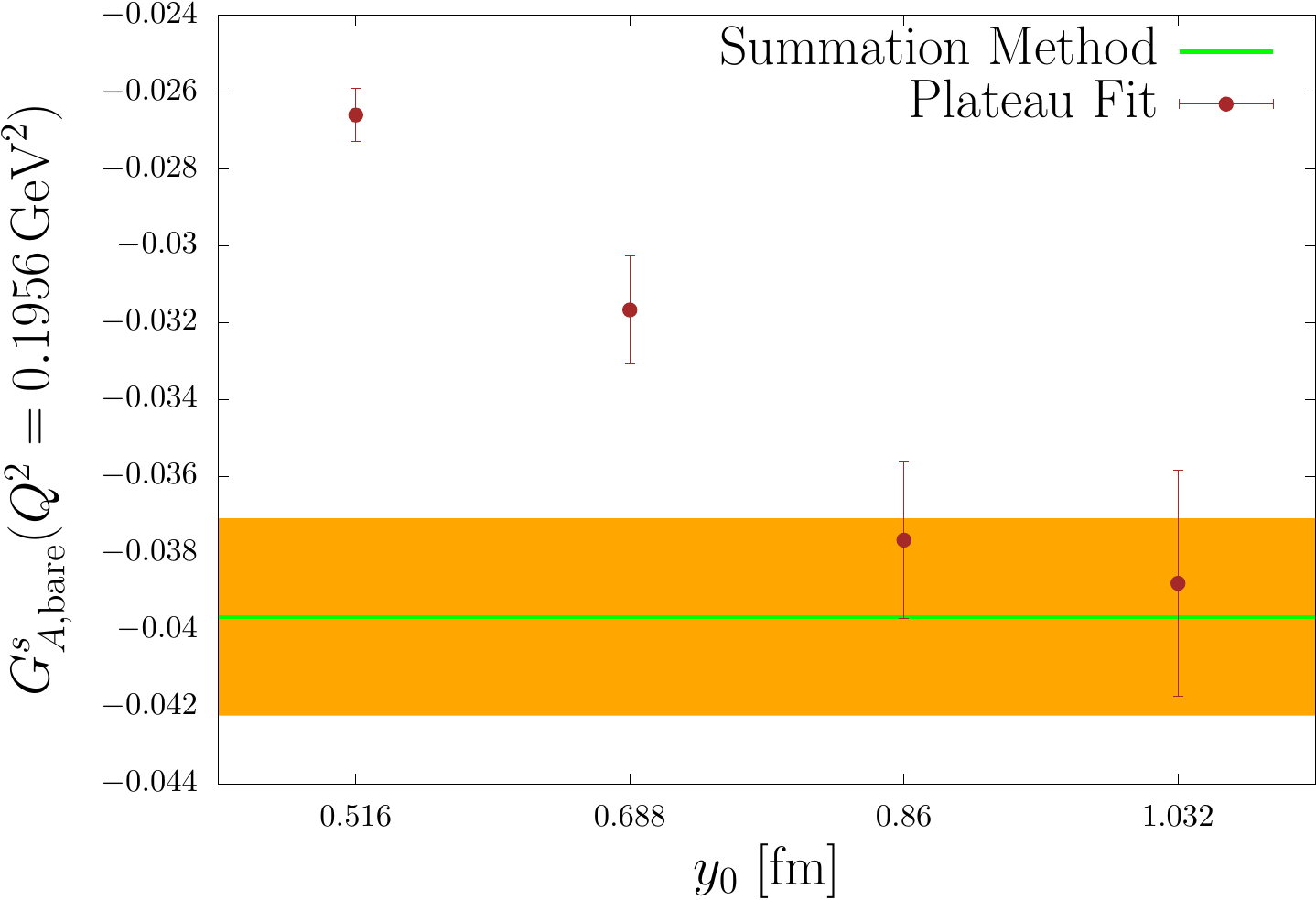}
\end{minipage}%
\begin{minipage}{\textwidth}
   \hspace{-8cm}
   \includegraphics[scale=0.4225]{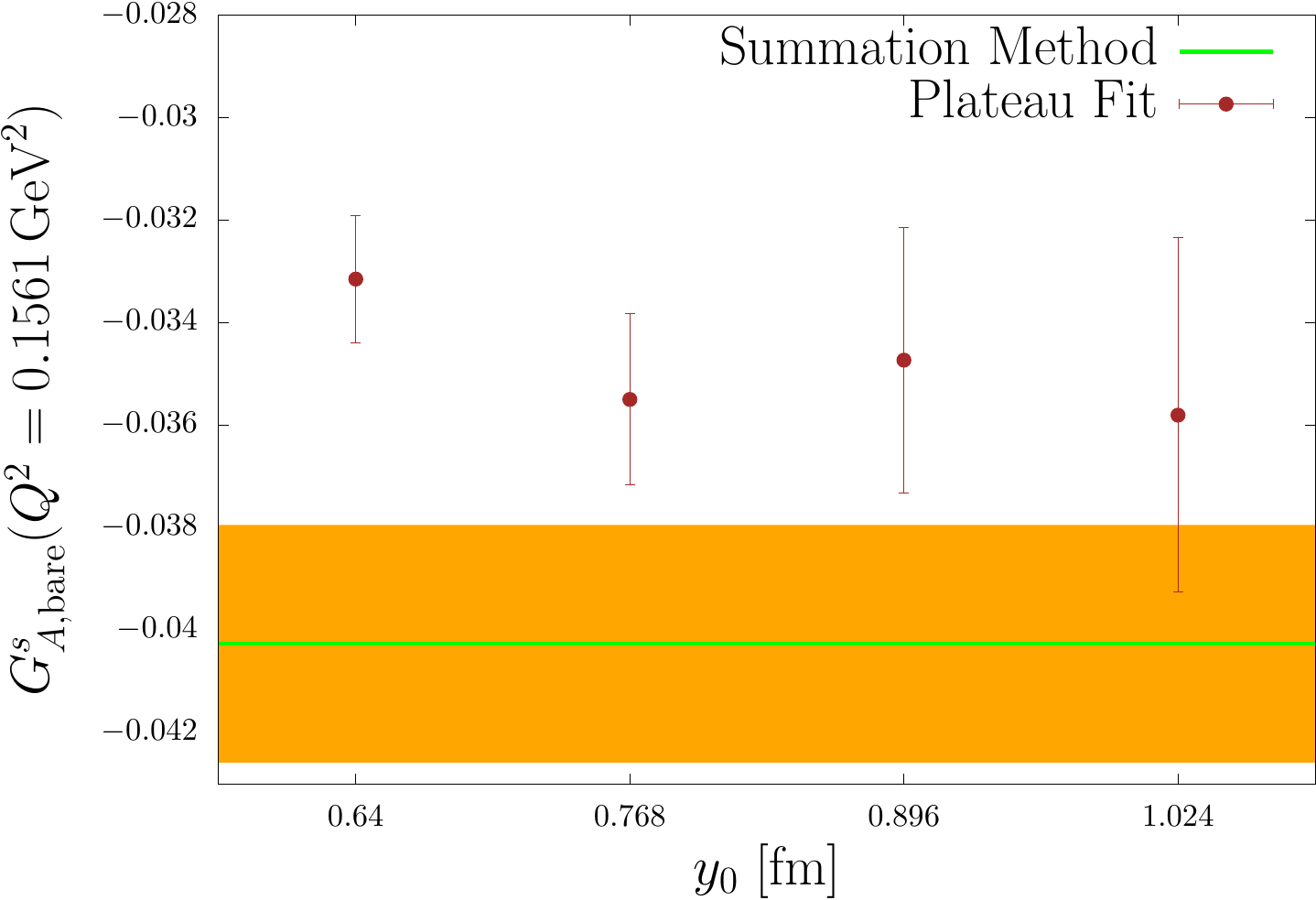}
\end{minipage}
\caption{Comparison of plateau fits at different source-sink separations $y_0$ and the summation method for two ensembles at $m_\pi=280\,\text{MeV}$ (left: $a=0.086\,\text{fm}$, right: $a=0.064\,\text{fm}$).}
\label{fig:exstate}
\end{figure}
\end{center}

\begin{center}
\begin{figure}[h]
\begin{minipage}{\textwidth}
   \includegraphics[scale=0.4225]{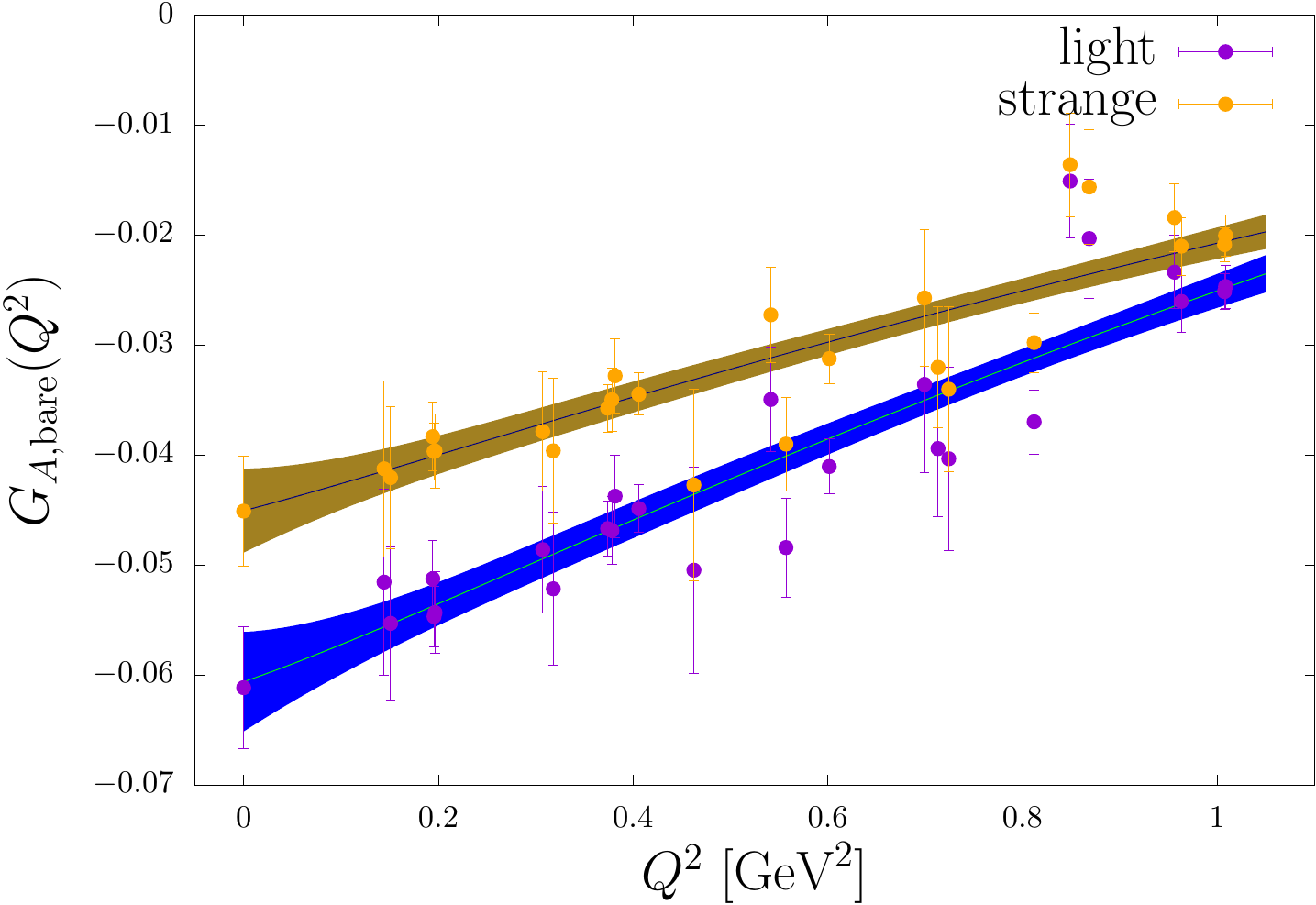}
\end{minipage}%
\begin{minipage}{\textwidth}
   \hspace{-8cm}
   \includegraphics[scale=0.4225]{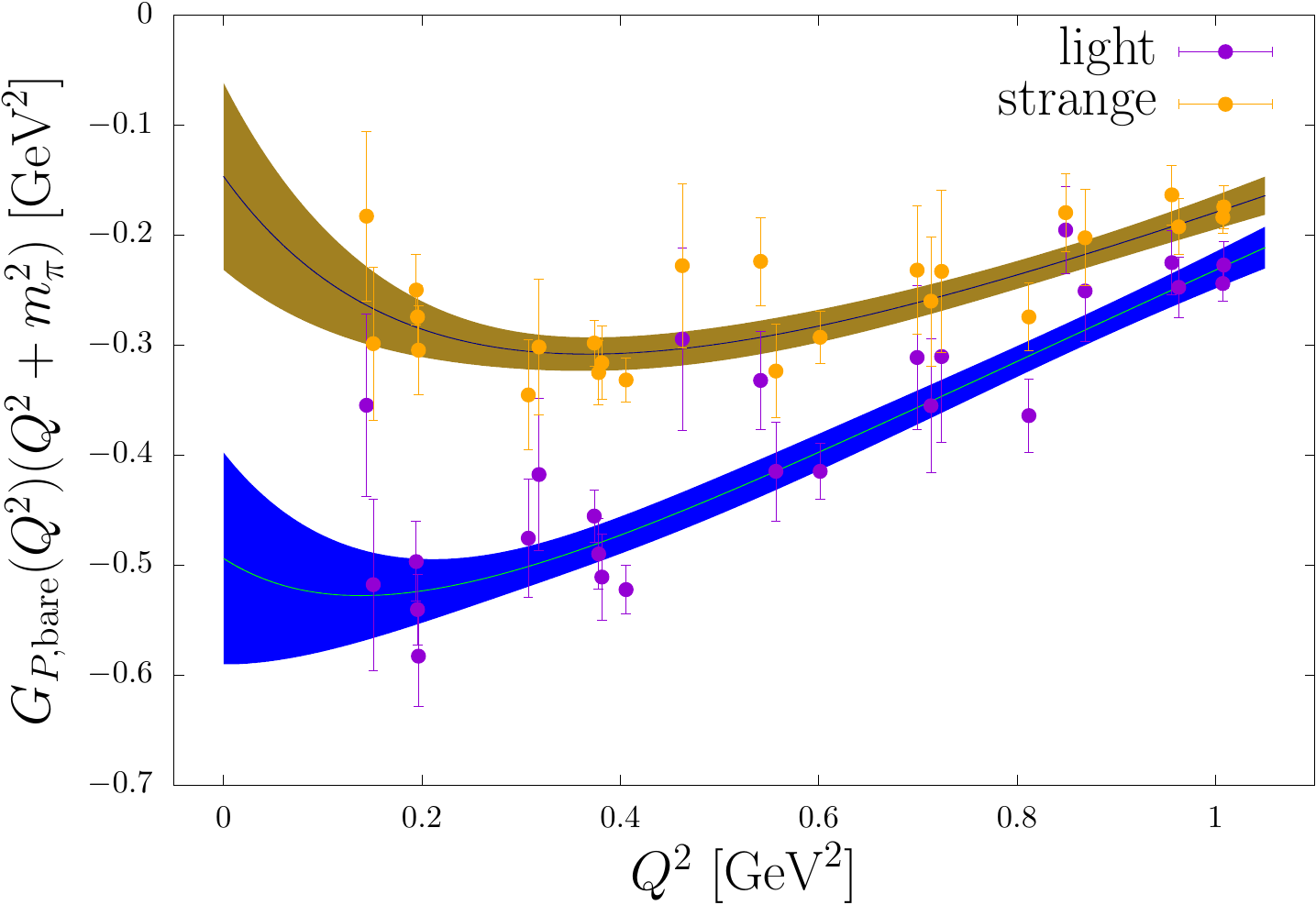}
\end{minipage}
\caption{The disconnected contribution of the light and strange quarks to the axial form factor (left) and the induced pseudoscalar form factor (right) for the ensemble with $a=0.086\,\text{fm}$ and $m_\pi=280\,\text{MeV}$.}
\label{fig:Q2deü}
\end{figure}
\end{center}

In \Fig{fig:exstate} we show the strange axial form factor for a particular non-vanishing $Q^2$ as a function of the source-sink separation $y_0$ used for the plateau fit and also include a band visualising the summation method result. For both ensembles excited-state contamination is visible but we find agreement between the plateau fits and the summation method at large enough $y_0$. In the following, we only consider the summation method results in order to have tidier plots. The $Q^2$-dependence of the disconnected axial vector form factors for the light and the strange quarks on the ensemble with $a=0.086\,\text{fm}$ and $m_\pi=280\,\text{MeV}$ are shown in \Fig{fig:Q2deü}. The results for the induced pseudoscalar form factor have been multiplied with $(Q^2+m_\pi^2)$ in order to remove the pion pole. The curves are z-expansion fits to fifth order with Gaussian priors for all coefficients $a_k$ with $k\geq 2$. Both the axial and the induced pseudoscalar form factor are found to be non-vanishing and negative.

\vspace{-1.7cm}

\begin{center}
\begin{figure}[h]
\begin{minipage}{\textwidth}
	\includegraphics[scale=0.4225]{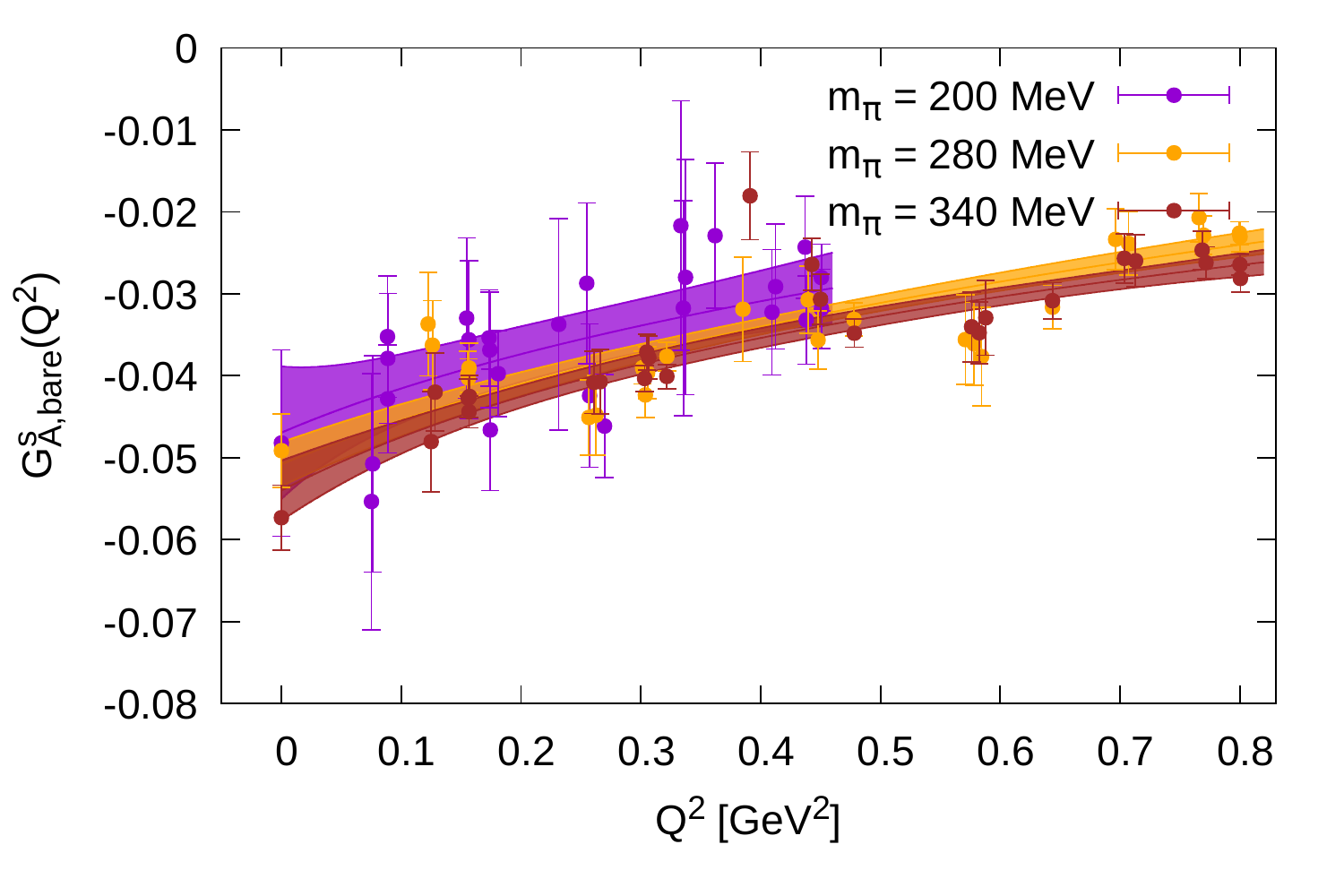}
\end{minipage}%
\begin{minipage}{\textwidth}
   \hspace{-8cm}
 	\includegraphics[scale=0.4225]{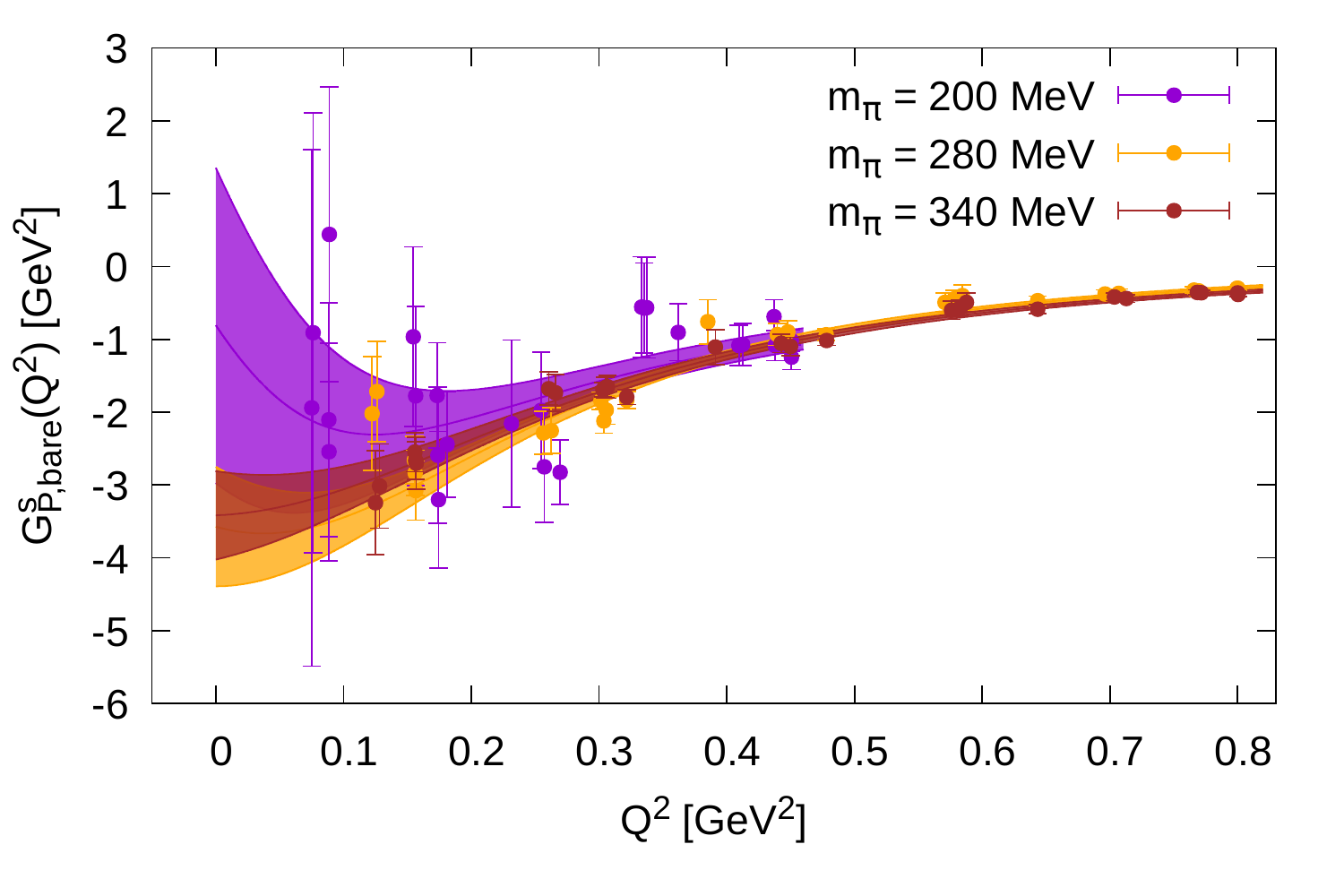}
\end{minipage}\\
\begin{minipage}{\textwidth}
	\includegraphics[scale=0.4225]{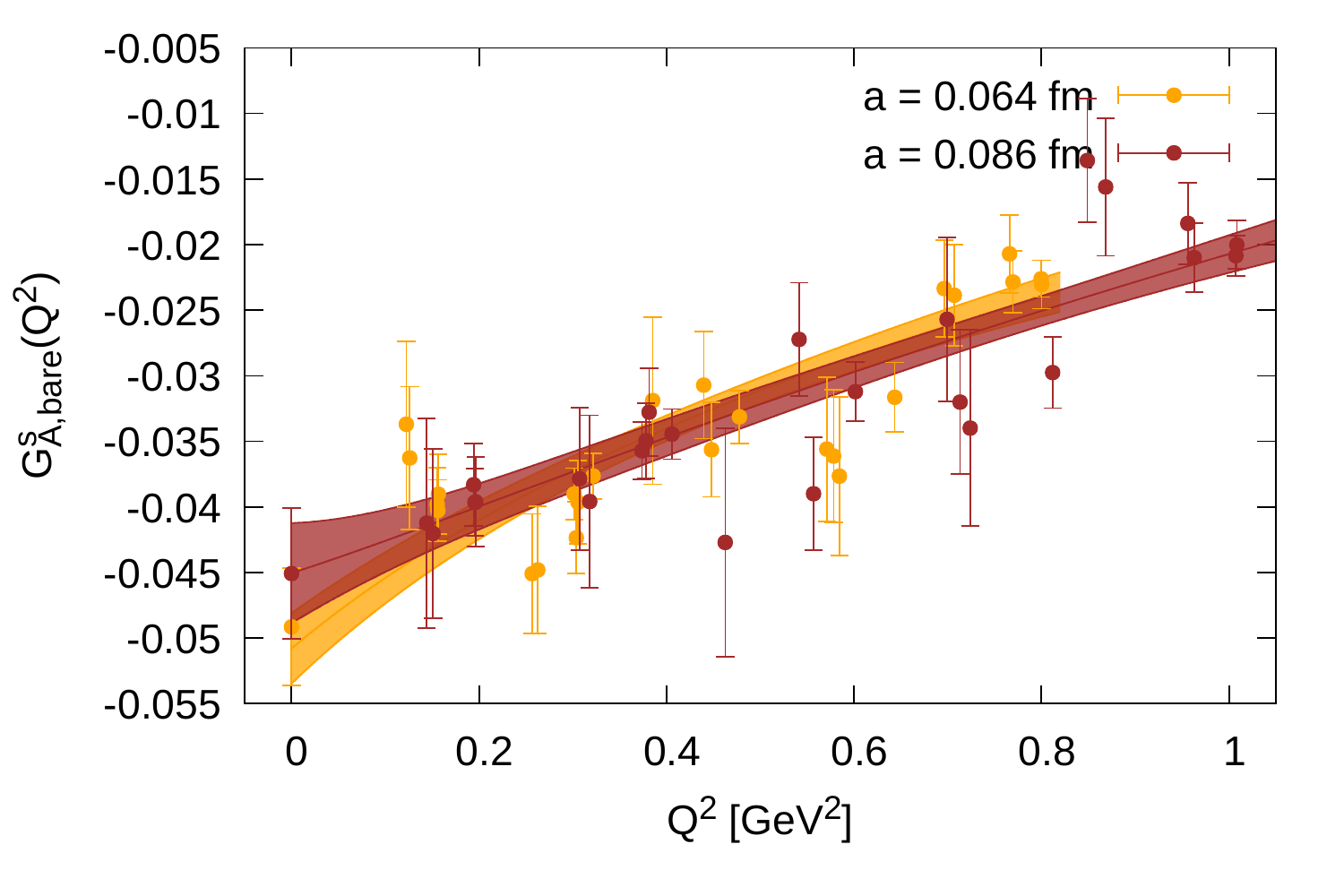}
\end{minipage}%
\begin{minipage}{\textwidth}
   \hspace{-8cm}
 	\includegraphics[scale=0.4225]{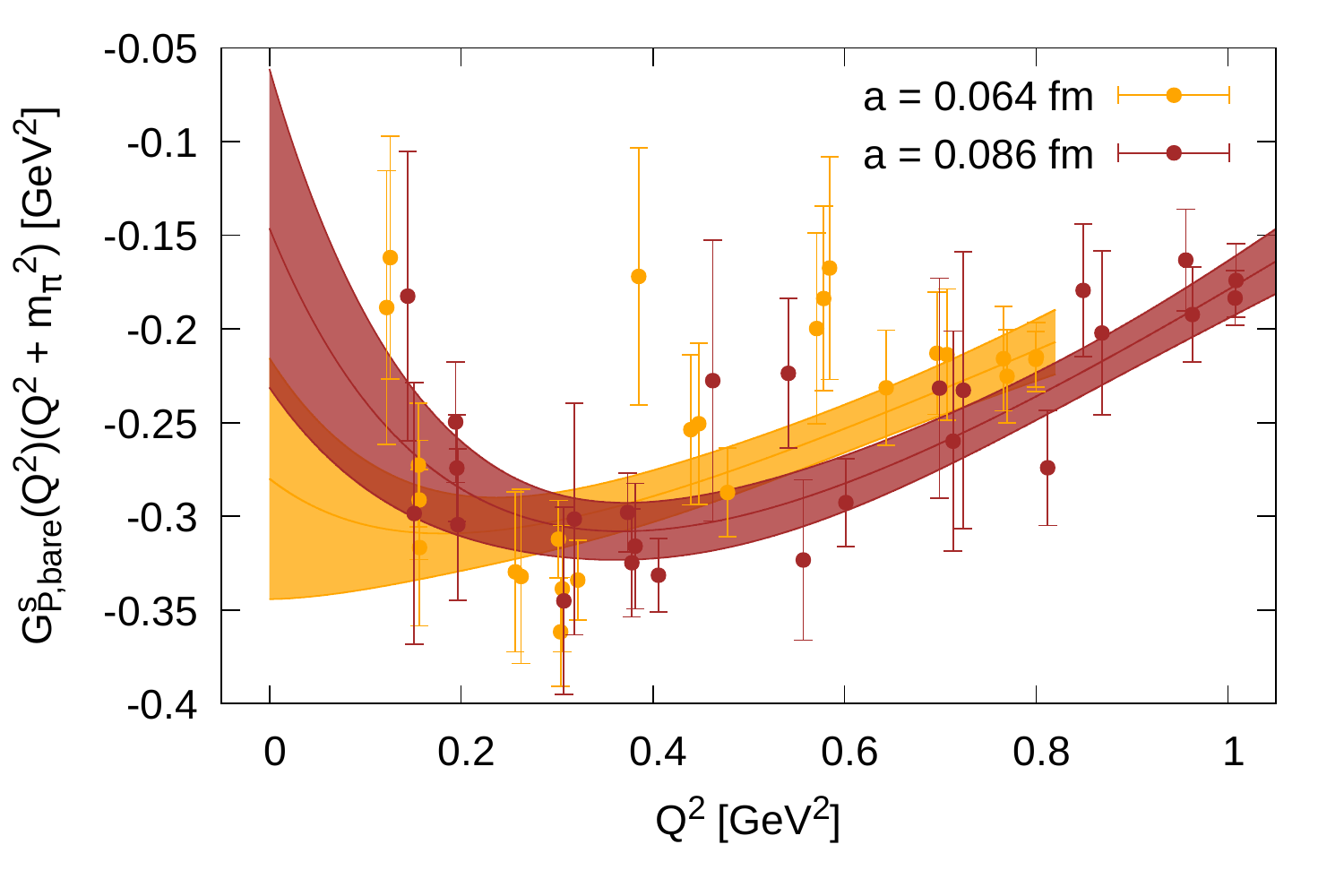}
\end{minipage}
\caption{Pion mass dependence at a lattice spacing of $a=0.064\,\text{fm}$ (top) and lattice spacing dependence at a pion mass of $m_\pi=280\,\text{MeV}$ (bottom) of the strange axial vector form factors.}
\label{fig:mpi_dep}
\end{figure}
\end{center}

\vspace{-0.8cm}
Lastly, the pion mass dependence at fixed lattice spacing and the lattice spacing dependence at fixed pion mass of the strange axial vector form factors is illustrated (\Fig{fig:mpi_dep}). At this level of statistics, we find the strange axial vector form factors to depend only mildly on the pion mass and the lattice spacing. In future work, we will include more ensembles into this analysis and attempt a continuum extrapolation. Furthermore, the investigation of disconnected contributions to the electromagnetic form factors is planned.

\vspace{-1.4cm}
\section*{Acknowledgements}
This research is supported by the DFG through the SFB 1044. K.O. is supported in part by DFG grant HI 2048/1-1. Calculations for this project were partly performed on the HPC clusters "{}Clover"{} and "{}HIMster II"{} at the Helmholtz-Institut Mainz and "{}Mogon II"{} at JGU Mainz. Additional computer time has been allocated through projects HMZ21 and HMZ36 on the BlueGene supercomputer system "{}JUQUEEN"{} at NIC, J\"ulich. Our programmes use the QDP++ library \cite{QDPpp} and deflated SAP+GCR solver from the openQCD package \cite{openQCD}, while the contractions have been explicitly checked using \cite{QCT}. We are grateful to our colleagues in the CLS initiative for sharing ensembles.

\end{document}